\documentclass[prl,twocolumn,showpacs,preprintnumbers,amsmath,amssymb]{revtex4}
\usepackage{graphicx}
\usepackage{dcolumn}
\usepackage{bm}
\usepackage{psfig}
\newcommand \be{\begin{eqnarray}}
\newcommand \ee{\end{eqnarray}}
\newcommand {\ket}[1]{|#1\rangle}
\newcommand {\bra}[1]{\langle #1|}
\begin{document}
\title{Correlated two-particle scattering on finite cavities}
\author{K. Morawetz$^{1,2}$, 
M. Schreiber$^1$,   B. Schmidt$^{1}$, A. Ficker$^{1}$ and P. Lipavsk\'y$^{3,4}$}
\affiliation{$^1$Institute of Physics, Chemnitz University of Technology, 
09107 Chemnitz, Germany}
\affiliation{
$^2$Max-Planck-Institute for the Physics of Complex
Systems, Noethnitzer Str. 38, 01187 Dresden, Germany}
\affiliation{$^3$Institute of Physics, Academy of Sciences, 
Cukrovarnick\'a 10, 16253 Prague 6, Czech Republic}
\affiliation{$^4$Faculty of Mathematics and Physics, Charles University,
Ke Karlovu 5, 12116 Prague 2, Czech Republic
}

\begin{abstract}
The correlated two-particle problem is solved analytically in the presence of a finite cavity. The method is demonstrated here in terms of exactly solvable models for both the cavity as well as the two-particle correlation where the two-particle potential is chosen in separable form. The two-particle phase shift is calculated and compared to the single-particle one. The two-particle bound state behavior is discussed and the influence of the cavity on the binding properties is calculated.
\end{abstract}
\maketitle

\section{Introduction}

Finite quantum structures are intensively studied concerning their transport properties, trapping behavior as well as considering them as artificial atoms \cite{FATHT02,CBHH02} which become especially interesting in reduced dimensions. The main tool to investigate the properties of such quantum cavities is the scattering of particles connected with excitations \cite{FCLCPDG02} or the current measurement \cite{RRAMMW88,FMMWMBBKW93,KLCSH02} through these structures. We want to regard as quantum cavities here also quantum dots or any finite structure. The effort of theoretical investigations has reached a fairly high level adequate to the obtained experimental quality of data. Nevertheless it is noticeable that almost all investigations are restricted to one-particle transport properties like current, conductance etc. On the other side the experimental abilities are on such level that also higher-order correlations and more than one-particle properties can be envisaged. 
\begin{figure}
\psfig{file=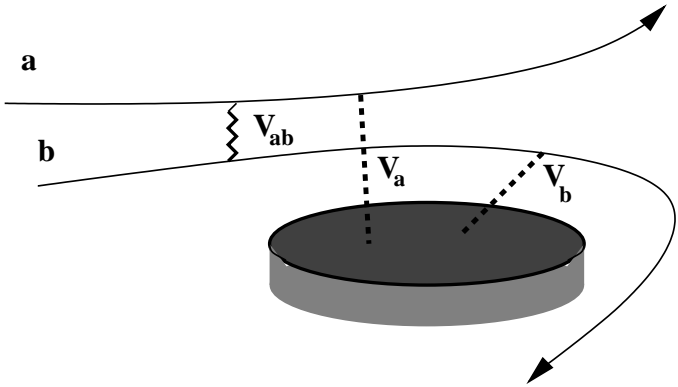,width=9cm}
\caption{Schematic picture of correlated two-particle scattering on a quantum cavity.\label{ill}}
\end{figure}
The aim of the present paper is to investigate two-particle correlations together with one-particle correlations on the same footing. Since these investigations suggest to look at a new type of experiments on quantum cavities we want to present an exactly solvable model for exploratory reasons. In this way the method is demonstrated which can then be used for any realistic nanostructure, with an amorphous or crystalline system as background provided the one-particle wave functions are available from e.g. standard ab-initio methods. 

As illustrated in figure~\ref{ill} we consider the gedanken experiment to scatter two particles with each other in the presence of a quantum cavity. In usual approaches it would be sufficient to know the scattering properties of single particles (though correlated) on the cavity in order to describe the transport properties. We want to demonstrate that the two-particle correlations in addition to the scattering on the quantum cavity lead to interesting new effects of coupled channels which might be worth an experimental investigation.  

As sketched by the figure, we face a two-center scattering which  is conveniently described by coupled-channel scattering theory \cite{T72}. A similar problem has been considered by Beregi et al. who solved the problem with the help of the Faddeev technique \cite{BLR70,BL71,Be72}. We will adopt here a simpler approach which is possible due to the Gell-Mann and Goldberger formulation \cite{GG53}. Due to the different occurring interactions, we define two channels, the one-particle scattering channel $V_1=V_a+V_b$ and the corresponding two-particle one, $V_2=V_{ab}$. The total ${\cal T}$-matrix can be found by the Gell-Mann and Goldberger two-potential  formula \cite{GG53} summarized in appendix~\ref{coupled},
\be
{\cal T}={\cal T}_1+(1+V_1 {\cal G}_1) {\cal T}_{ab} (1+{\cal G}_1 V_1).
\label{GG}
\ee
Here the one-particle ${\cal T}$-matrix is given by 
\be
{\cal T}_1=V_1+V_1 {\cal G}_1 V_1
\ee
and the correlated two-particle ${\cal T}$-matrix ${\cal T}_{ab}$ is a solution of the modified Lippmann-Schwinger equation 
\be
{\cal T}_{ab}=V_2+V_2 {\cal G}_1 {\cal T}_{ab}
\label{ab}
\ee
where the full solution of the one-particle channel enters in ${\cal G}_1$.
For that one assumes that the one-particle problem is solved and the propagator in the one-particle channel is known
${\cal G}_1=(\omega -{\cal H}_1)^{-1}$.
This scheme is advisable if one channel can be solved exactly and the other channel is then described by a Lippmann-Schwinger equation (\ref{ab}) with the help of the solved channel. This method has been compared to other coupled-channel methods \cite{G74} and successfully applied to defects in potentials \cite{CR91} and in developing a hybrid representation of the nuclear potential \cite{RC90}. We will use this scheme here and will employ an exactly solvable model for both channels.

\section{Model}

\subsection{Model cavity}
 For the cavity we choose a model of an opaque wall in three dimensions which can be found in text books \cite{F94}.
This model is able to simulate properties of opaque confining traps as well as parametric resonances. The potential in the single-particle channel takes the form
\be
{V}_a(r)={1\over 2 m} {\Omega \over R} \delta (r-R)
\label{pote}
\ee
with the reduced mass $m$ of particles $a$ and $b$ scaled out. Thus we have an opaque delta potential at the radial distance $r=R$ from the center of cavity. The wave functions out- and inside the cavity decouple completely in the limit of hard spheres, $\Omega \to \infty$, where the interior of the cavity is characterized by  discrete energy levels. For the opaque wall, $\Omega < \infty$, outside and inside are coupled and resonances occur at the virtual levels.
 
\begin{figure}
\psfig{file=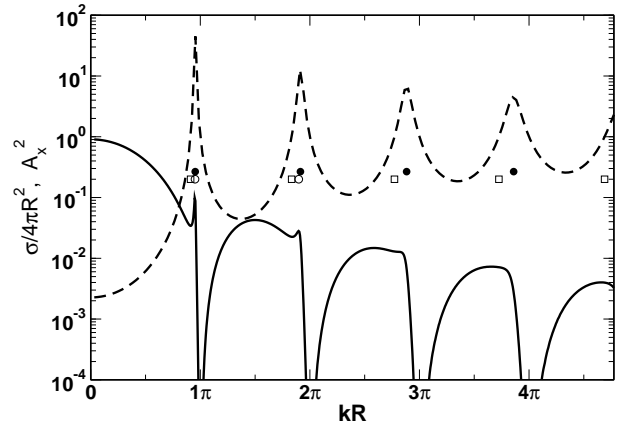,width=9cm,angle=0}
\caption{The single-particle cross section on the cavity (\protect\ref{sigma1}) with zeros at $kR=n\pi$ (solid line) and the amplitude (\protect\ref{A1}) (dashed line) versus momenta for $\Omega=20$. The filled circles marks the maxima of the amplitude (\protect\ref{xnres0}), the open circles the maxima of the cross section according to (\protect\ref{xnres}) and the open squares the minima and maxima of the cross section according to (\protect\ref{xnres1}).   \label{a2fig}}
\end{figure}

The wave function of this radial problem has the form $\Psi_n(r)=\bra{r}n\rangle=\chi_n(r)/r$. While the subscript $n$ denotes general quantum numbers the specific wave function here shall be characterized by continuous momenta $n=k$ such that
\be
\chi_k(r)=c\left \{ \begin{array}{ll}A_{kR} \sin{k r} &; r<R \cr \sin{(k r+\phi)} &;r>R 
\end{array} \right ..
\label{opa}
\ee
The one-particle phase shift $\phi$ is determined by
\be
\cot{(x +\phi)}-\cot{x}=\Omega/x,
\label{ph0}
\ee
where $x=k R$. The amplitude can be written as
\be
A_x^2={\sin^2{(x+\phi)}\over \sin^2{x}}={1+\tan^2{x}\over \tan^2{x}+(1+{\Omega \over x}\tan{x})^2}
\label{A1}
\ee
and shows resonances at the virtual levels
\be
\tan 2 x_\nu=-2 {x_\nu/ \Omega},
\label{xnres0}
\ee
with even $\nu$ which approach
the levels $x_\nu=k_\nu R=\nu \pi$ in case of an 
impenetrable wall, $\Omega\to\infty$.
Assuming only s-wave scattering the corresponding single-particle cross section reads
\be
\sigma(k)={4 \pi \over k^2}{1\over  1+\cot^2{\phi}}
\label{sigma1}
\ee
which has zeros at $x_n=n\pi$. From (\ref{ph0})
one finds that the cross section has a limited number of maxima at
\be
-2{x_\mu\over \Omega}=\sin{2 x_\mu}
\label{xnres}
\ee
depending on the potential strength, $\Omega-\pi/2\ge 2 \pi |\mu|$. These maxima are accompanied by minima and an infinite number of further maxima at
\be
-2{x_\mu\over \Omega}=\tan{2 x_\mu}.
\label{xnres1}
\ee
These features are illustrated in figure~\ref{a2fig}.

\subsection{Two-particle problem with separable potential}

We study in this paper the combined problem of single- and two-particle scattering and will use a separable representation of the two-particle interaction which provides tremendous technical simplifications.  One can show that the 
effect of any finite-range potential on the scattering wave function $\Psi$ can be represented in separable form \cite{Ko92,KPML97},
\be
{\cal V}\ket{\psi}=\sum\limits_{ij}\ket{g_i}\tau_{ij}\bra{g_j}p\rangle,
\ee
where $\ket{p}$ is the undisturbed plane wave and $g_i$ are form factors.
This is known as the expansion of the wave function inside the potential range and outlined in appendix~\ref{proof}. Separable potentials are powerful tools for constructing unknown nuclear matter potentials from available experimental scattering phase shifts by inverse scattering theory \cite{KK97}. The application of such separable potentials is versatile ranging from plasma physics \cite{SJR89} to medium modification of transport properties in nuclear matter \cite{SRS90,ARS94,MOR94} up to high field problems in superconductivity \cite{Mtc01}. 

According to the Gell-Mann and Goldberger formula (\ref{GG}) we assume the one-particle wave function $\ket{n_1}$ to be known
\be
{\cal H}_1 \ket {n_1}=E_{n_1} \ket{n_1}.
\label{single}
\ee
The two-particle ${\cal T}$-matrix equation (\ref{ab}) is now represented in momentum states. We note the momentum  difference of the two particles by small letters, e.g. $p$, and the center-of-mass momentum by capital letters, e.g. $P$. Introducing a complete set $|n_1n_2\rangle$ according to (\ref{single})  we obtain
\be
&&\bra{p_1P}{\cal T}_{ab}\ket{p_2P'}=V_{p_1p_2}\delta_{PP'}
\nonumber\\
&&\qquad+\sum\limits_{n_1 n_2 p p'\bar{P}}V_{p_1 p}{\langle pP |n_{1}n_{2}\rangle \langle n_{1}n_{2}| p'\bar{P}\rangle \over \omega-E_{n_1n_2}+i\eta} \bra{p'\bar{P}}{\cal T}_{ab}\ket{p_2 P'}.\nonumber\\&&
\label{tmat}
\ee 
By convention we use $\sum\limits_q=\int {d{\bf q}\over (2 \pi)^3}$ and set $\hbar=1$ within this paper.
Equation (\ref{tmat}) is the exact formulation of the quantum two-particle problem in channel 2. 
It can be solved with the help of such a separable potential \cite{Y59}
\be
V_{pq}=\lambda g_p g_q
\label{sep}
\ee
with form factors $g_p$ as given in appendix~\ref{app}. While generally any finite-range potential can be represented in a separable form of finite rank, we restrict ourselves here to rank-one potentials for exploratory reasons. The technique remains the same for high-rank potentials.

\section{Two-particle scattering on finite cavity}

\subsection{Free two-particle problem}\label{free2}
The case without cavity or unrestricted two-particle scattering is solved by plane waves with the center-of-mass momentum $K$ and the difference momentum $k$ as good quantum numbers, $|n_1n_2\rangle=|kK\rangle$. Performing the $k,K$ integrations, the Lippmann-Schwinger equation (\ref{tmat}) reads with $\bra{p_1P}{\cal T}_{ab}\ket{p_2P'}=\delta_{PP'}t_{p_1p_2}(P)$
\be
t_{p_1p_2}(P)=\lambda g_{p_1}g_{p_2}+\lambda \sum\limits_q g_{p_1} g_{q} {t_{q p_2}\over \omega-{P^2\over 4 m}-{q^2\over m}+i \eta}
\label{inf0}
\ee
which is easily solved as 
\be
\label{Jinf}
t_{p_1p_2}(P)&=&{\lambda g_{p_1}g_{p_2} \over 1-\lambda J(P,\omega)}\nonumber\\
J(P,\omega)&=&\sum\limits_q {g_q^2\over \omega-{P^2\over 4 m}-{q^2\over m}+i \eta}.
\ee 
This result is given explicitly for different form factors in appendix~\ref{app}.

\subsection{Two-particle scattering and cavity}

Now we are going to consider the disturbed two-particle scattering problem, which should reduce to
(\ref{Jinf}) in the limit of unrestricted two-particle scattering.
The equation for the ${\cal T}$-matrix (\ref{tmat}) can be simplified by 
introducing the modified form factors
\be
G_{n_1 n_2}(P)&=&\sum\limits_p g_p \langle pP|n_1 n_2\rangle.
\label{G}
\ee
Splitting off the trivial factors
\be
\bra{p_1P}{\cal T}_{ab}\ket{p_2P'}=\lambda g_{p_1}g_{p_2} t(PP'),
\label{fact}
\ee
from (\ref{tmat}) the equation for the center-of-mass dependence
follows
\be
&&t(PP')=\delta_{PP'}+\lambda \sum\limits_{n_1n_2} {G_{n_1n_2}(P)\sum\limits_{\bar P}G_{n_1n_2}^*({\bar P})t({\bar P}P') \over \omega -E_{n_1n_2}+i \eta}.
\nonumber\\&&
\label{cent}
\ee
The completeness of $\ket{n_1n_2}$ implies the relation
\be
\sum\limits_{n_1n_2}G_{n_1n_2}(P)G^*_{n_1n_2}(P')=\delta_{PP'} \sum\limits_q g_q^2.
\label{comp}
\ee
We introduce the auxiliary function
\be
h^*_{n_1n_2}(P')=\sum\limits_{\bar P}G_{n_1n_2}^*({\bar P})t({\bar P}P')
\label{aux}
\ee
from which the ${\cal T}$-matrix can be obtained with the help of (\ref{comp}) as
\be
t(PP')&=&{1\over \sum\limits_q g_q^2}\sum\limits_{n_1n_2} G_{n_1n_1}(P) h^*_{n_1n_2}(P').
\label{th}
\ee
Multiplying (\ref{cent}) with $G_{m_1m_2}^*({P})$ and integrating over $P$ we obtain an equation for the auxiliary function (\ref{aux})
\be
&&h^*_{m_1m_2}(P')=G_{m_1m_2}^*({P'})+\sum\limits_{n_1n_2} {H_{m_1m_2n_1n_2} h^*_{n_1n_2}(P') }
\nonumber\\&&
\label{sys}
\ee
with
\be
H_{m_1m_2n_1n_2}=\lambda {\sum\limits_{P}G_{m_1m_2}^*({P})G_{n_1n_2}(P)\over \omega -E_{n_1n_2}+i \eta}.
\ee

For cavities possessing only discrete states in which case $n_i$ and $m_i$ are natural numbers, (\ref{sys}) is a linear equation system which can be easily solved. With the help of (\ref{th}) and (\ref{fact}) the ${\cal T}$-matrix is consequently found exactly. 

The situation becomes more complicated if continuum states are present in the cavity. Then the quantum numbers $\bra{n_1n_2}$ becomes continuous variables like in the case of the free two-particle problem and (\ref{sys}) turns into an integral equation. We cannot present at the moment the analytical exact solution of the combined problem consisting of bound and scattering states. The numerical scheme consists in discretizing this continous integral and solving the linear equation system (\ref{sys}). This complete numerical solution will be devoted to a forthcoming work. To get a first physical insight, we will instead restrict ourselves in this paper 
only to cavities with no bound states, i.e. only scattering states, and moreover to scattering events where the center-of-mass momenta of the two particles are not remarkably changed.

\subsection{Approximated solution}
Due to the spatial structure of the one-particle problem, in general we will not have translational symmetry. This means that the two-particle problem need not to have the same center-of-mass momenta of incoming and outgoing particles. Only in special cases like the zero center-of-mass momentum and no change of the total momentum due to release or capturing of momentum by the cavity, we can assume approximately a diagonal form $t(PP')\approx \delta_{PP'}t(P)$. 

We understand this approximation as considering the effect of the cavity on the free two-particle problem (\ref{inf0}) and (\ref{Jinf}) in first order. Such a diagonal solution is obtained, if for the momentum integral over ${\bar P}$ on the right hand side of (\ref{cent}) we apply the mean value theorem of integrals and approximate the mean value ${\tilde P}$ by the center-of-mass momentum $P$
\be
\sum\limits_{\bar P}G_{n_1n_2}^*({\bar P})t({\bar P}P')&=&t({\tilde P}P')\sum\limits_{\bar P}G_{n_1n_2}^*({\bar P})\nonumber\\
&\approx& t({P}P')\sum\limits_{\bar P}G_{n_1n_2}^*({\bar P}).
\ee
Then (\ref{cent}) is easily solved
and the diagonal ${\cal T}$-matrix is 
\be
\bra{p_1P}{\cal T}_{ab}\ket{p_2 P'}=\delta_{PP'} {\lambda g_{p_1}g_{p_2}\over 1-\lambda J(P,\omega)}
\label{sol}
\ee
with 
\be
J(P',\omega)&=&
\sum\limits_{n_1n_2} {\sum\limits_PG_{n_1n_2}(P) G^*_{n_1 n_2} (P')\over \omega -E_{n_1n_2}+i\eta }.
\label{J}
\ee

Alternatively we can obtain this result by assuming {\em ad-hoc} a diagonal ${\cal T}$-matrix $t(PP')\approx \delta_{PP'} t(P)$ which according to (\ref{cent}) fulfills the integral equation 
\be
&&t(P')\delta_{PP'}=\nonumber\\
&&\qquad\delta_{PP'}
+\lambda \sum\limits_{n_1n_2} {G_{n_1n_2}(P)\sum\limits_{\bar P}G_{n_1n_2}^*({\bar P})t(P')\delta_{\bar{P}P'}\over \omega -E_{n_1n_2}+i \eta}.
\nonumber\\
&&
\label{1}
\ee
Integrating this equation over $P$ we obtain the same approximate 
solution (\ref{sol})
\be
t(P')
&=&{1\over 1-\lambda J(P',\omega)}.
\label{2}
\ee

It is interesting to note here that (\ref{2}) is also a solution of a nontrivial integral equation. To see this one can integrate  (\ref{1}) over $P'$ instead over $P$ as done above. One then obtains the integral equation
\be
t(P)&=&1
+\lambda \sum\limits_{n_1n_2} {G_{n_1n_2}(P)\sum\limits_{\bar P}G_{n_1n_2}^*({\bar P})t({\bar P})\over \omega -E_{n_1n_2}+i \eta}.
\label{int}
\ee
Therefore we can conclude that (\ref{2}) is also an approximate diagonal solution of the integral equation (\ref{int}) which has not been known previously to the authors' knowledge.

The approximate solution (\ref{sol}) together with (\ref{J}) and (\ref{G}) is the main result which we will employ within this paper. It represents a generalization of the standard separable ${\cal T}$-matrix result, e.g. of Ref. \cite{Y59}, in two respects: (i) It describes the two-particle scattering on a finite cavity or on an arbitrary structure represented by the one-particle wave functions and (ii) it gives the exact solution of the two-particle quantum problem in the case of conservation of total momentum before and after scattering.

Before continuing  we want to show first how the limit of unrestricted two-particle scattering is reached. Unrestricted scattering means the absence of the cavity so that the single-particle wave functions become plane waves, $\ket{n_1n_2}=\ket{kK}$ as discussed in section~\ref{free2}.
Consequently, (\ref{G}) takes the form
\be
G_{kK}^{\rm inf}(P)=g_{k} \delta_{PK}
\label{liminf}
\ee
and (\ref{J}) is
\be
J^{\rm inf}(P\omega)=\sum\limits_q g_q^2 {1\over \omega -E_{{P\over 2}+q}-E_{{P\over 2}-q}+i\eta}
\label{infinit}
\ee
which is the standard expression for unrestricted two-particle scattering (\ref{Jinf}) and separable interaction. In nuclear physics this case is labelled infinite matter.

\section{Analytical results for two-particle scattering on an opaque wall}

We are now going to evaluate the result (\ref{sol}) with (\ref{J}) and (\ref{G}). Restricting the calculation to a rank-one separable potential characterized by only one form factor, e.g.  $g_p=\exp{(-p^2/4\beta^2)}/\beta^2$, and the potential strength $\lambda$, one can describe two properties of the two-particle correlations, the scattering length $a_0$ and the range of the potential $r_0$ as illustrated in the appendix \ref{app}. For clarity we will first concentrate only on the scattering length and will separate the effects of potential range. Therefore it is advisable to start with the contact potential which is represented by the limit of zero range $\beta \to \infty$. In a second step we will discuss the effects of finite range of the potential.

\subsection{Contact potential for two-particle correlations}

The separable potential allows us to adopt the contact potential by a special limit using a parameter-dependent coupling constant $\lambda(\beta)$ such that the scattering length is kept fixed. In this way one ensures intrinsically the necessary renormalizations which otherwise have to be performed by subtracting diverging terms to fix the scattering length to the physical value \cite{LL42}. 

Mathematically, the limit $\beta \to \infty$ with fixed scattering length is realized if we rewrite the coupling constant $\lambda$ in terms of the scattering length. This is obtained from the general relation between phase shift $\varphi$ and the ${\cal T}$-matrix
\be
\tan \varphi=\left . {{\rm Im} T\over {\rm Re} T} \right |_{\omega={p^2\over m}}=    \left . {\lambda J^{\rm inf}_{\rm I}\over 1-\lambda J^{\rm inf}_{\rm R}}\right |
\label{contact}
\ee
where we will abbreviate the real and imaginary parts of the unrestricted two-particle scattering expression (\ref{infinit}) as $J^{\rm inf}_{\rm R}={\rm Re} J^{\rm inf}(P,{p^2\over m})$ and $J^{\rm inf}_{\rm I}={\rm Im} J^{\rm inf}(P,{p^2\over m})$, respectively.
One knows that for contact potentials \cite{aocomm}
\be
\tan \varphi^{\rm inf}=p a_0
\label{phinf}
\ee
for all momenta $p$ and we express the parameter-dependent coupling constant $\lambda(\beta)$ from (\ref{contact}) as
\be
\lambda={p a_0 \over J^{\rm inf}_{\rm I}+p a_0 J^{\rm inf}_{\rm R}}.
\ee
Inserting this expression into (\ref{sol}) we obtain
\be
&&{\cal T}_{ab}(P,\omega)={\lambda g_{p_1}g_{p_2}\over 1-\lambda J(P,\omega)}={p a_0 g_{p_1}g_{p_2}\over J^{\rm inf}_{\rm I}-p a_0  (J^{\rm inf}_{\rm R} -J(P,\omega))}.
\nonumber\\&&
\label{tabo}
\ee
Now we are ready to perform the limit of zero range $\beta \to \infty$. From (\ref{J}) and (\ref{infinit}) one sees that the form factors $g_p=1/(\beta^2+p^2)$ appear with the same order in the numerator as well as in the denominator which cancel in the limit of zero range. Consequently, (\ref{tabo}) for the ${\cal T}$-matrix (\ref{sol}) approaches 
\be
&&{\cal T}_{ab}(P,\omega) \to {\cal T}_{ab}^{0}(P,\omega)= {p a_0 \over J^{0{\rm inf}}_{\rm I}-p a_0  (J^{0{\rm inf}}_{\rm R} - J^0(P\omega))}
\nonumber\\&&
\label{solc}
\ee
where the additional superscript $0$ denotes that both expressions (\ref{J}) and (\ref{infinit}) have to be taken with $g_p\to 1$. The difference $J^{0{\rm inf}}_{\rm R}-J^0$ in the denominator (\ref{solc}) ensures the convergence of the expression since ${\rm Re} J$ and $J^{0{\rm inf}}_{\rm R}$ diverge. 

It is instructive to check the ${\cal T}$-matrix for contact potentials without cavity. From (\ref{infinit}) one has 
\be
J^{0{\rm inf}}_{\rm I}(P,\omega+{P^2\over 4 m})&=&-{m\over 4 \pi} \sqrt{m \omega} \Theta(\omega)
\nonumber\\
J^{0{\rm inf}}_{\rm R}(P,\omega+{P^2\over 4 m})&=&{1\over 2 \pi^2}\int\limits_0^\infty {dk k^2 \over \omega -{k^2\over m}}
\label{J0}
\ee 
which leads to the ${\cal T}$-matrix (\ref{solc}) for contact potentials in unrestricted two-particle scattering
\be
&&{1\over {\cal T}^{0{\rm inf}}\left (P,\omega+{P^2\over 4 m}\right )}=
{m\over 4 \pi} \left \{
\begin{array}{ll}
{\!-\!{1\over a_0}+i p} &; \omega={p^2\over m}\cr&\cr
{\!-\!{1\over a_0}-\sqrt{m E_b}} &;\omega=-E_b
\end{array}
\right ..
\nonumber\\&&
\ee 
Though (\ref{J0}) diverges, the solution (\ref{solc}) remains finite. With the help of (\protect\ref{contact}) one can convince oneself that (\protect\ref{phinf}) is fulfilled for positive energies $\omega=p^2/m$ as it should be \cite{signcomm} and that in the case that the scattering length is negative indicating attraction, the bound state $E_b=1/ma_0^2$ represents the pole of the ${\cal T}$-matrix at $\omega=-E_b$.

Now we are going to calculate $J^0$ according to (\ref{J}) including the finite cavity represented by the single-particle wave function $|n\rangle$ in (\ref{opa}).  The explicit form for (\ref{G}) reads
\be
G_{n_1 n_2}^0(P)&=&\sum\limits_p \langle pP|n_1n_2\rangle
=\int d {\bf r} {\rm e}^{-i {\bf r P}} \Psi_{n_1}({\bf r}) \Psi_{n_2}({\bf r})
\nonumber\\&&
\ee
where the spatial representation of the one-particle wave function is $\Psi_{n_1}({\bf r})=\langle {\bf r}|n_1\rangle$.
This means that 
\be
\sum\limits_P G^0_{n_1 n_2}(P)&=&\Psi_{n_1}(0)\Psi_{n_2}(0)
\nonumber\\
G^0_{n_1 n_2}(0)&=&\sqrt{2} \delta_{n_1n_2}.
\label{help}
\ee
The last equality follows from the completeness of the set of single-particle wave functions as well as the normalization 
\be
\int d{\bf r} |\Psi_{n_1}|^2=\sqrt{2}
\label{norm}
\ee 
ensuring that the normalization of the wave function $|n_1 n_2\rangle$ is 2 for the two particles. 

Now we can give a simple expression for (\ref{J}) in the case of center-of-mass scattering $P=0$ which we want to discuss in the following. We obtain with the help of (\ref{help})
\be
J^0(0,\omega)=\sum\limits_{n_1} {|\Psi_{n_1}(0)|^2 \sqrt{2}\over \omega -2 E_{n_1}+i\eta}.
\label{form}
\ee
Together with (\ref{J0}) we have obtained the two-particle ${\cal T}$-matrix for contact interaction (\ref{solc}) provided the one-particle wave function of the cavity $\Psi_{n_1}({r})=\langle {r}|n_1\rangle=\chi_{n_1}(r)/r$ and the single-particle energy $E_{n_1}$ according to (\ref{single}) are known. 
For this purpose we use (\ref{opa}) of the opaque wall and the necessary part of the wave function in (\ref{form}) reads
\be
\Psi_{n_1}(0)=c k A_{kR}
\ee
with the amplitude given by (\ref{A1}).
We determine the normalization $c$ in such a way that the particle should be in a sphere with radius $L$ which is set to infinity in the end. For such large boundaries we can employ radial cyclic boundary conditions, $\chi(r+L)=\chi(r)$, such that we have
\be
\sum\limits_{n_1}={L\over 2 \pi}\int\limits_0^\infty dk.
\label{suml}
\ee
Calculating the norm (\ref{norm}) with (\ref{opa}) it follows that
\be
\lim\limits_{L\to\infty} {L c^2\over 2 \pi}={\sqrt{2}\over 4 \pi^2}.
\label{liml}
\ee
Therefore, the arbitrary large fictitious boundary length $L$ is dropping out in (\ref{form}) and we obtain 
\be
{\rm Re} J^0(0,\omega)&=&{1\over 2 \pi^2}\int\limits_0^\infty dk {k^2 A_{kR}^2 \over \omega -{k^2\over m}}
\nonumber\\
{\rm Im} J^0(0,\omega={p^2\over m})&=&-{m p\over 4 \pi} A_{pR}^2.
\label{J1}
\ee
This allows us to determine the ${\cal T}$-matrix (\ref{solc}) with the help of (\ref{J0}) as
\be
{m\over 4 \pi a_0} {\cal T}_{ab}^{0}(0,{p^2\over m})=\left (-1-{a_0 \over R} F_\Omega(p R)+i p a_0 A_{pR}^2 \right )^{-1}
\label{T1}
\ee
with
\be
F_\Omega(z)={2\over \pi}\int\limits_0^\infty dx {x^2\over z^2-x^2}\left (A_x^2-1\right ).
\label{F}
\ee
Please note that the $-1$ term comes from the subtraction of (\ref{J0}) in (\ref{solc}) which renders the expression finite. The function $F$ can be given analytically as presented in appendix~\ref{Fana}. From (\ref{T1}) we obtain the two-particle phase shift
\be
\tan{\varphi}(p)={p R \,\,A_{p R}^2\over F_\Omega(p R)+\frac{R}{a_0}}.
\label{phi1}
\ee

\begin{figure}
\psfig{file=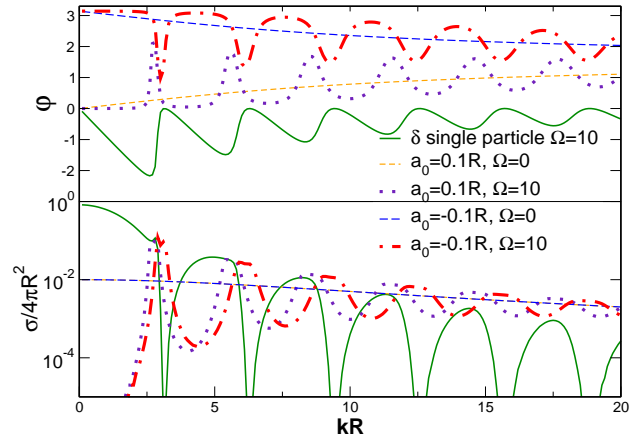,width=9cm,angle=0}
\caption{The one-particle phase shift (top) of an opaque wall (\protect\ref{ph0}) versus wave vector and the free two-particle phase shift (\protect\ref{phinf}) as well as the combined ones (\protect\ref{phi1}). The wall parameter is $\Omega=10$ which corresponds to $F_\Omega(0)=0.91$ and the two-particle scattering length $a_0=\pm0.1R$. In the bottom figure the corresponding cross sections (\protect\ref{sigmaa}) are plotted.  \label{10_0.1}}
\end{figure}

In figure~\ref{10_0.1} we plot the phase shifts for a repulsive opaque wall with $\Omega=10$ and a weak two-particle scattering length of $a_0=\pm0.1R$. The one-particle phase shift $\phi$ is negative indicating repulsion and has the typical oscillatory structure \cite{F94} due to the resonances at the virtual levels. The free two-particle phase shift $\varphi^{\rm inf}$ of (\ref{phinf}) approaches $\pi/2$ for large $k$ and starts from zero for positive $a_0$ and from $\pi$ for negative scattering length indicating a bound state according to the Levinson theorem \cite{GW64}. 
The two-particle phase shift (\ref{phi1}) for the case including the opaque wall follows the free two-particle scattering case for such small scattering lengths though the oscillations of the one-particle problem are visible. Interestingly these oscillations are shifted slightly when positive and negative scattering lengths are compared. 

\begin{figure}
\psfig{file=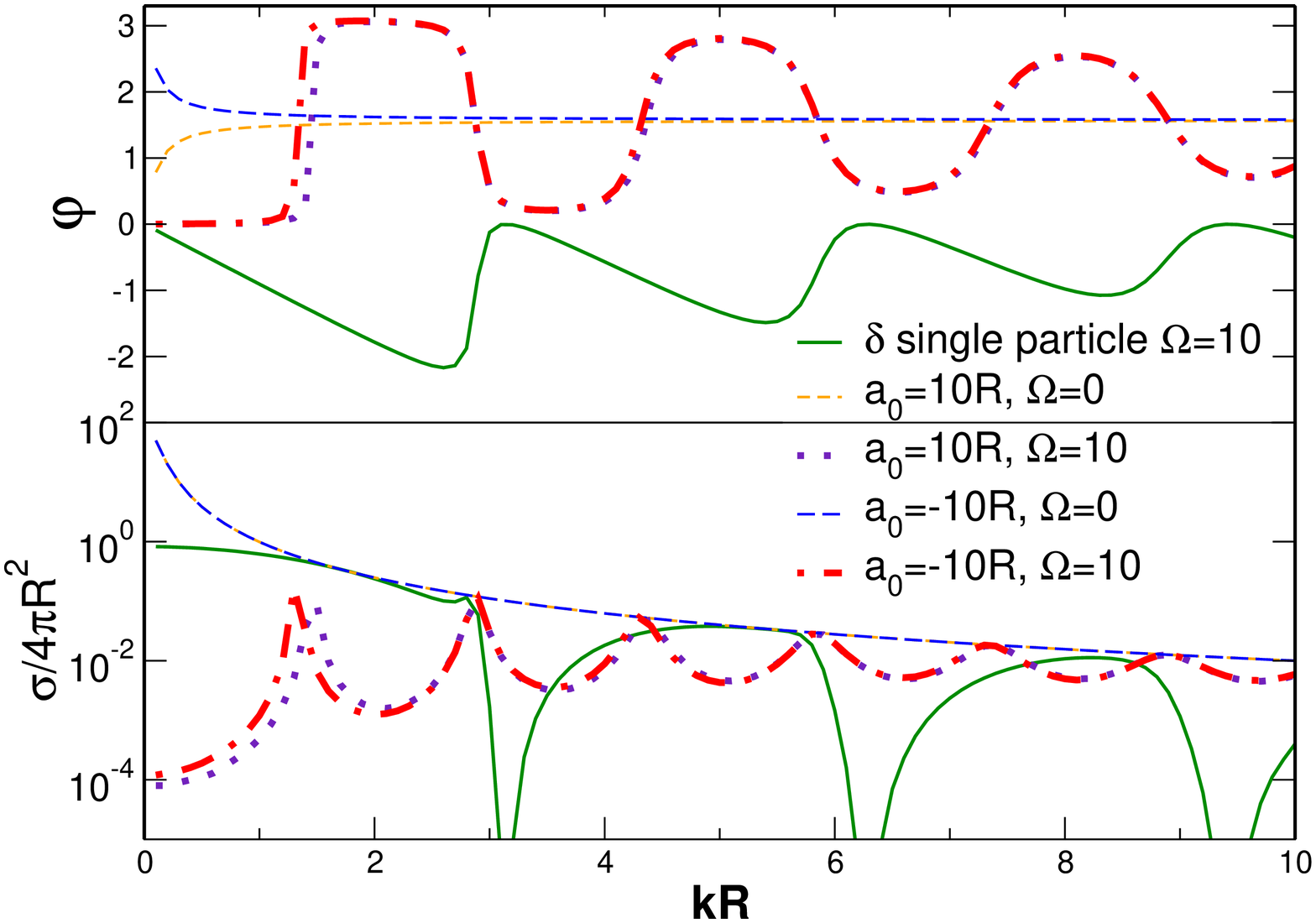,width=9cm,angle=0}
\caption{Same as figure~\ref{10_0.1} but for $a_0=\pm 10R$.  \label{10_10}}
\end{figure}

With the help of (\ref{phi1}) the corresponding cross section takes the form 
\be
\sigma&=&{4 \pi \over p^2} \sin^2{\varphi}
={4 \pi R^2} {A_x^4\over x^2 A_x^4+(F_\Omega(x)+\frac{R}{a_0})^2}
\label{sigmaa}
\ee
which will be determined by the structure of the one-particle amplitude $A$ and a combination of $A$ and $F$ dependent on $a_0$ in the denominator of (\ref{sigmaa}). The latter leads to additional structure in the cross section and can be understood as an interference between the one-particle and two-particle properties. This can barely be recognized in figure~\ref{10_0.1}  where a second maximum appears besides the oscillations given by the one-particle cross section. The appearance of this second maximum is much more pronounced when we go to larger scattering lengths corresponding to stronger two-particle correlation as in figure~\ref{10_10}. Therefore, it is a correlation effect in the two-particle channel.

\begin{figure}
\psfig{file=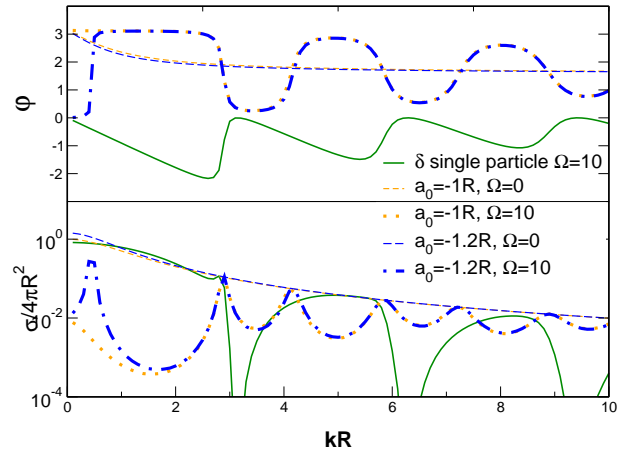,width=9cm,angle=0}
\caption{Same as figure~\protect\ref{10_0.1} but for $a_0=-1.2R$ and $a_0=-R$.  \label{10_-1}}
\end{figure}

Figure~\ref{10_10} also shows that the two-particle problem in the presence of a cavity has no bound state anymore though the free two-particle problem has one. The repulsive opaque wall has destroyed the bound state formation. 
In order to understand this behavior we consider the ${\cal T}$-matrix not at positive energies $\omega=k^2/m$ as done so far but at negative energies where the pole yields the bound state energy $\omega=-E_b$. The ${\cal T}$-matrix (\ref{solc}) reads in this case
\be
&&{\cal T}_{ab}^0(0,-E_b)
={{-4 \pi a_0/m} \over 1+a_0\sqrt{mE_b}+{a_0 \over R} F_\Omega(i R \sqrt{m E_b}) }
\ee
with $F$ from (\ref{F}). Consequently, the bound state energies are given by the pole
\be
-\frac{R}{a_0}=F_\Omega(ib)+b
\label{bcond1}
\ee
with $b=R \sqrt{m E_b}$. It is easy to prove that 
$F_\Omega(ib)+b\ge F_\Omega(0)$, the explicit analytical expression can be found in appendix~\ref{Fana}. This means,
as soon as the cavity parameter $\Omega$, $R$ and the two-particle scattering length $a_0$ fulfill the inequality 
\be
-\frac{R}{a_0} \ge F_\Omega(0)
\label{bcond}
\ee
we will have two-particle bound states. In figure~\ref{10_0.1} this is the case for negative scattering lengths since $R/a_0=-10$ and $F_{10}(0)=10/11$ while no bound states occur for positive scattering length $R/a_0=10$. Oppositely, in figure~\ref{10_10} for both positive and negative scattering lengths we have  $R/a_0=\pm 0.1< F_{10}(0)=10/11$ and no bound states appear anymore. In this figure the free two-particle problem has a bound state for negative scattering length while the repulsive wall allows no bound state for these parameters.
According to (\ref{bcond}) the limiting scattering length is $a_0=-1.1 R$ for $\Omega=10$. In figure~\ref{10_-1} we plot the cases just slightly above and below that critical value. One sees that the bound state present in the free case is destroyed by the cavity if the scattering length is more negative than the lower critical value.

It is now interesting to consider the attractive wall $\Omega<0$. Then we might expect that bound states are favored in any case which is surprisingly not true.

\begin{figure}
\psfig{file=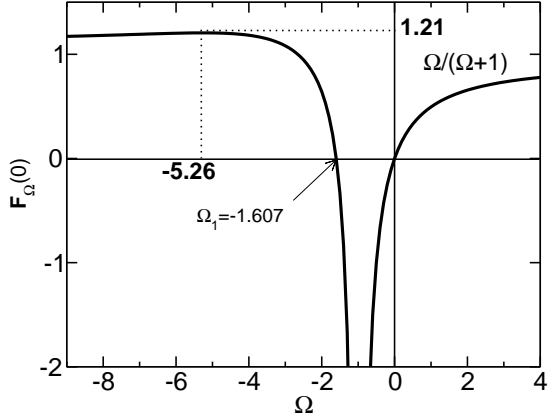,width=9cm,angle=0}
\caption{The function $F_\Omega(0)$ of (\protect\ref{F}) which allows bound states if $-R/a_0\ge F_\Omega(0)$ according to (\protect\ref{bcond}).\label{fb}}
\end{figure}

We plot the function $F_\Omega(0)$ in figure~\ref{fb}. 
In the range  $\Omega_1<\Omega<0$, $F_\Omega(0)$ becomes negative. Therefore any negative $a_0$ leads to a bound state according to (\ref{bcond}). Both two-particle scattering as well as the opaque wall are attractive and also their combination acts attractively. Positive $a_0$ will lead to a bound state if $a_0\ge -R/F_\Omega(0)\ge 0$. Though the wall is attractive to both particles, bound states occur only if the two-particle correlations are larger than a critical value.

\begin{figure}[h]
\psfig{file=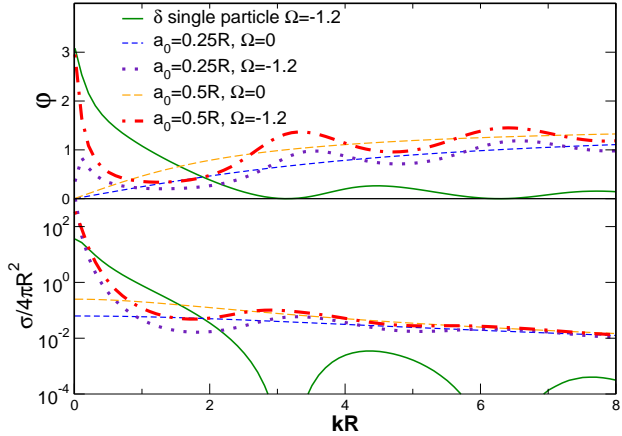,width=9cm,angle=0}
\caption{Same as figure~\ref{10_0.1} but with $\Omega=-1.2$ which corresponds to $F_\Omega(0)=-3.33$ and $a_0=0.25R$ and $a_0=0.5R$. \label{-1.2_25}}
\end{figure}

This is illustrated in figure~\ref{-1.2_25}.
In this figure we have chosen free two-particle scattering parameters which do not lead to bound states. The attractive opaque wall creates a bound state for $a_0=0.5 R$ while no bound state occurs for $a_0=0.25 R$.

For the range $\Omega<\Omega_1$ the function $F_\Omega(0)$ is positive. This means that the same discussion holds as for positive $\Omega$. According to (\ref{bcond}) $a_0$ must be negative and $a_0 \ge -R/F_\Omega(0)$. This means that bound states occur only if the two-particle scattering length is larger than a critical value. 
The appearance of such critical values for the two-particle scattering length is an unexpected effect of interference. 
Please note also that if $1\le F_\Omega(0) \le 1.21$ there are two values of $\Omega$ which lead to the same $F_\Omega(0)$. This means we can have two different cavities $\Omega$ which yield the same bound state.

\subsection{Finite range effects\label{finite}}

After discussing the effect of the scattering length we want to return to the more realistic situation of finite-range two-particle interaction. Therefore we choose a separable form of the potential and evaluate (\ref{J}) explicitly. For exploratory reasons we choose the exponential form factor of appendix~\ref{expo} and the center-of-mass system $P=0$.

Introducing the dimensionless integrals
\be
&&G_{n_1n_2}(0)\!=\!\int \!{d {\bf q}\over (2 \pi)^3} g_q \Psi_{{\bf k}_1}({\bf q})\Psi_{{\bf k}_2}(-{\bf q})\equiv {8 R c^2\over \beta^2} \tilde G_0({x}_1,{x}_2)\nonumber\\
&&\int {d {\bf p}\over (2 \pi)^3} G_{n_1n_2}(p)
={32 \beta R c^2\over \sqrt{\pi}} {\bar G}(x_1,x_2)
\label{inte}
\ee
with $x_n=R k_n$, we obtain with the help of (\ref{suml}) and (\ref{liml}) 
\be
\lambda J(0,\omega^*)\!=\!{2^6 \xi \beta^2 R^2\over \pi^{5/2}} \!\int \limits_0^\infty \! dx_1 dx_2 {\tilde G_0(x_1,x_2) {\bar G}(x_1,x_2)\over \omega^*-x_1^2-x_2^2+i \eta}.
\label{js}
\ee
Here we used $\omega^*=2 m R^2\omega$ and the dimensionless coupling constant $\xi$ which can be written in terms of the scattering length $a_0$ according to (\ref{aoro})
\be
\xi={m\lambda \over 4 \pi^2 \beta^3}=-{1\over \sqrt{2 \pi}+{\pi\over \beta a_0}}.
\ee 
Consequently, the potential is attractive for scattering length $a_0>0$ or $a_0<-\sqrt{\pi/2}/\beta$ and repulsive otherwise. The potential range according to (\ref{exp}) follows from (\ref{aoro}) as
\be
r_0={1\over \beta} \left (2 \sqrt{2\over \pi}+{1\over a_0 \beta}\right )
\ee
which shows that for negative scattering length we should have $a_0<-\sqrt{\pi/2}/2\beta$ to render the potential range positive.

\begin{figure}
\psfig{file=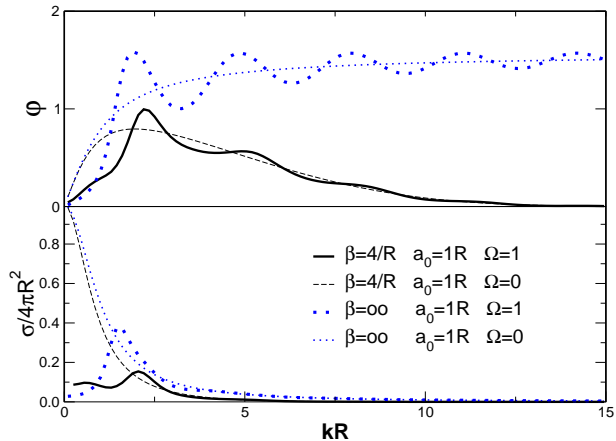,width=9cm}
\caption{The two-particle phase shift (top) and cross section (bottom) in the presence of the cavity (thick lines) and without cavity (thin lines). The contact-potential case $\beta\to\infty$ (dotted lines) is compared with the finite-range case (solid and dashed lines). The scattering length is $a_0=R$.}\label{boc4_1_1}
\end{figure}

It is convenient to use the spectral representation following from (\ref{js})
\be
{\rm Re}\lambda J(0,\omega^*)\!&=&\frac{1}{\pi} \int\limits_0^\infty d\omega {{\rm Im} \lambda J(0,\omega)\over \omega^*-\omega}\nonumber\\
{\rm Im}\lambda J(0,\omega^*)\!&=&\!-{2^5 \xi \beta^2 R^2\over \pi^{3/2}} 
\nonumber\\&\times&
\!\!\int \limits_0^{\sqrt{\omega^*}}\! dx_1 \!\left . {\tilde G_0(x_1,\sqrt{\omega^*\!-\!x_1^2}) {\bar G}(x_1,\sqrt{\omega^*\!-\!x_1^2})\over \sqrt{\omega^*\!-\!x_1^2}}\right .
.\nonumber\\&&
\label{js1}
\ee 
The explicit analytical integrals $\tilde G_0$ and ${\bar G}$ can be found in appendix~\ref{integrals}.

\begin{figure}
\psfig{file=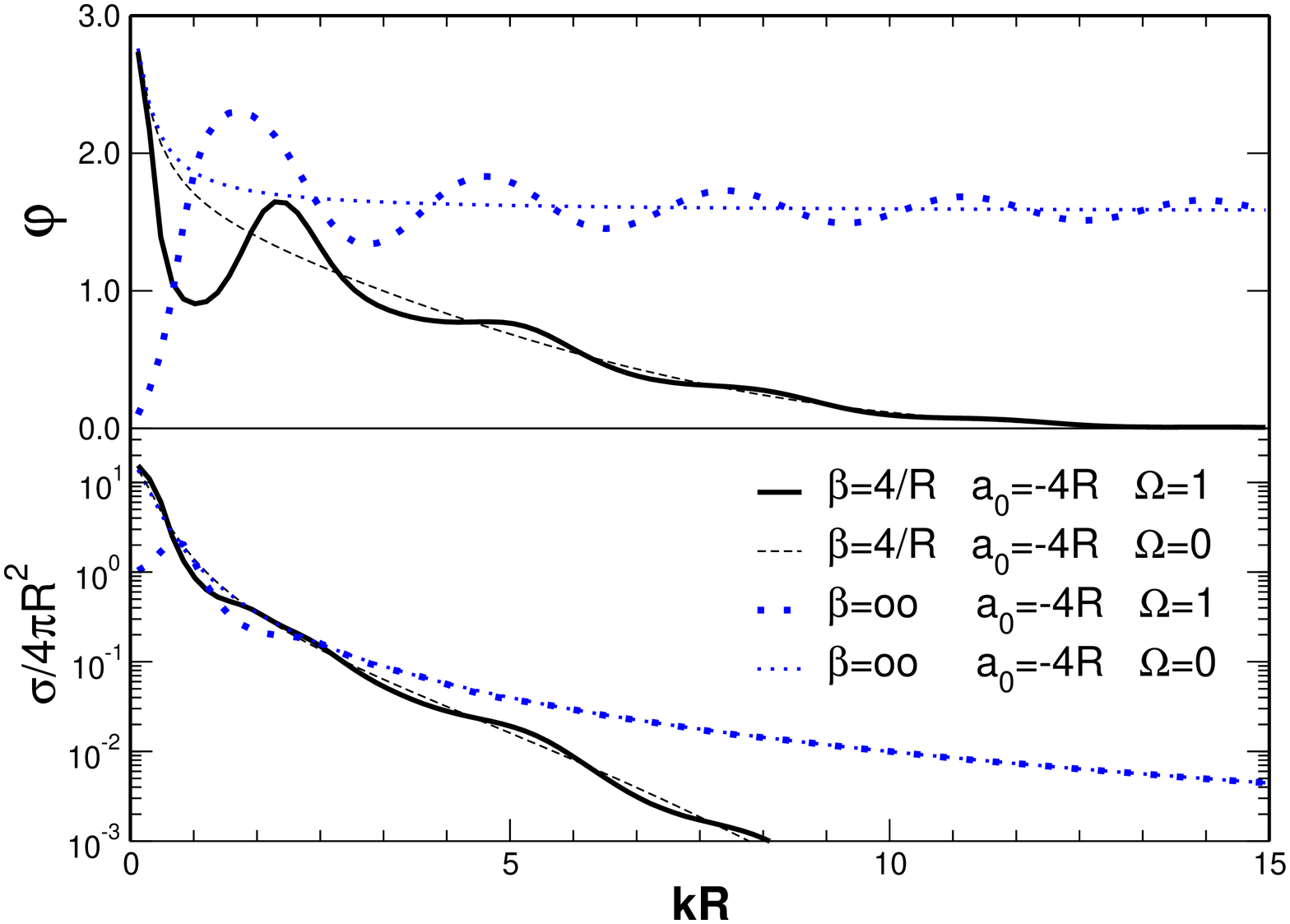,width=9cm}
\caption{Same as figure~\protect\ref{boc4_1_1}, but for $a_0=-4 R$.}\label{boc4_1_-4}
\end{figure}

\begin{figure}
\psfig{file=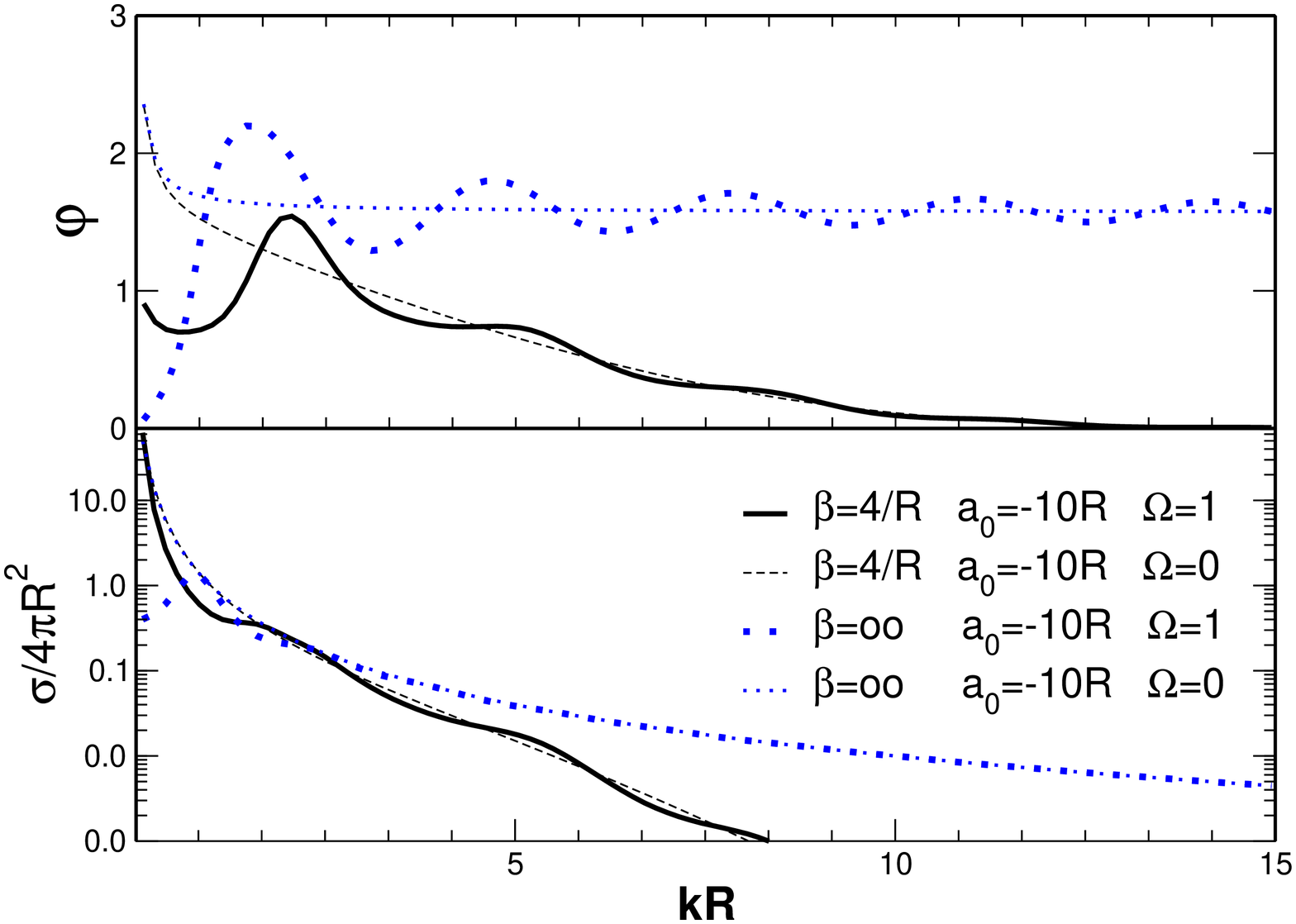,width=9cm}
\caption{Same as figure~\protect\ref{boc4_1_1}, but for $a_0=-10 R$.}\label{boc4_1_-10}
\end{figure}

The results are presented in figures~\ref{boc4_1_1}-\ref{boc4_1_-10}.
We compare the case with a cavity to the free two-particle scattering. The limit of contact potential $\beta \to \infty$ of the last chapter is plotted for comparison too. In figure~\ref{boc4_1_1} we have chosen a scattering length which does not allow the formation of bound states. The two-particle phase shift follows the general trend of the phase shift without cavity but shows characteristic oscillations which are visible in the cross section too. In the figure~\ref{boc4_1_-4} we chose such a scattering length that the free two-particle scattering has a bound state. While for the contact potential the repulsive opaque wall already destroys the bound state, the bound state survives in the finite-range potential. This means that the finite range enhances the correlations.

Decreasing the potential strength of the attractive two-particle scattering by a more negative scattering length the repulsive opaque wall will destroy also the bound state for the finite-range potential. In figure~\ref{boc4_1_-10} we plot the limiting case where the bound state is just still realized. 

From the figures it is obvious that the finite-range potentials are more realistic than the contact potentials. For higher energies the phase shifts for the contact potential approach an artificial value of $\pi/2$ while the phase shifts for the finite-range potential approach zero as they should. For finite-range potentials also the cross section falls off faster than for contact potentials.

\section{Summary}

We have investigated the coupled-channel problem of two particles interacting with themselves and with a cavity. We have chosen exactly solvable models for both the cavity as well as the two-particle potential in order to demonstrate the method. With the help of the single-particle wave function of the cavity, the two-particle problem is solved. For cavities with bound states the ${\cal T}$-matrix is determined exactly. In the case of additional scattering states we restrict ourselves only to processes where the change of the center-of-mass momenta are negligible and present an approximate solution. 

The influence of the cavity on the two-particle properties like scattering phase shift and bound states is investigated. We observe oscillations in the single-particle and two-particle phase shifts. While the oscillations in the single-particle phase shift are the standard resonances at the virtual levels of the cavity, the two-particle phase shift shows besides these oscillations also interferences which are visible in a splitting of the maxima in the cross section.

For repulsive opaque walls the scattering length must be negative but larger than a critical value in order that bound states remain. For attractive walls there is a specific range of negative wall parameters for which any negative scattering length will lead to bound states since both the cavity wall and the scattering length act attractively. In addition also positive scattering lengths will lead to bound states in this case provided the scattering length is larger than a critical value. For more attractive walls we could show that bound states are only possible for negative scattering lengths which are, however, limited by a negative critical value. The occurrence of such lower critical values is an unexpected effect of interference.  

The model discussed here has the merit to be analytically solvable in all parts and serves as a toy model for more realistic situations. In particular, the finite cavity can be easily described by a more realistic wave function which enters the two-particle problem in such a way as worked out here. We intended to provide in this way a clear method which is also easily applicable to the case of two-particle scattering on any finite or more realistic structure, like crystals or amorphous structures. In this case one has to provide the single-particle wave functions e.g. by an ab-initio method and can use the method here to calculate the two-particle correlations explicitly.

The discussion here has been limited to the two-particle correlations for exploratory reasons. In a forthcoming work we will show how this method is applicable for the situation of many-particle correlations.

\acknowledgements
This work has been supported by DAAD. The discussions with Ingrid Rotter
which have been of great value are gratefully acknowledged.

\appendix

\section{Formulas from coupled-channel scattering theory}\label{coupled}
\subsection{Many coupled channels}
We consider a system consisting of $i$ channels.
Using the free propagator
\be {\cal G}_0(\omega) = {1 \over \omega - {\cal H}_0 + i \eta}
\ee
and the propagator including the interaction of the $i$th channel
\be {\cal G}_i(\omega) = {1 \over \omega - {\cal H}_i + i \eta},
\ee
the Lippmann-Schwinger equation for the $i$th-channel propagator reads
\be
{\cal G}_i(\omega) = {\cal G}_0(\omega)+{\cal G}_i(\omega) V_i {\cal G}_0(\omega)
\label{calG}
\ee
corresponding to the $i$th-channel ${\cal T}$-matrix
\be
{\cal T}_i(\omega)=V_i+V_i {\cal G}_i(\omega) V_i.
\ee

The total ${\cal T}$-matrix is described by the Lippmann-Schwinger equation with respect to the total potential $V=\sum\limits_i V_i$ and can be written as a sum of auxiliary ${\cal T}$-matrices
\be
{\cal T}(\omega)=\sum\limits_i V_i + \sum\limits_i V_i {\cal G}_0(\omega) {\cal T}(\omega) = \sum\limits_i {\cal T}'_i(\omega).
\label{sumT}
\ee
These auxiliary ${\cal T}$-matrices are defined by
\be
{\cal T}'_i&=&V_i+V_i {\cal G}_0 {\cal T}
\label{defT}
\ee
and  can be rewritten as
\be
{\cal T}'_i&=&V_i+V_i {\cal G}_i{\cal T}-V_i{\cal G}_iV_i{\cal G}_0 {\cal T}
\nonumber\\&=&
V_i+V_i {\cal G}_iV_i\sum\limits_{j\ne i}V_i{\cal G}_iV_j(1+{\cal G}_0 {\cal T})
\nonumber\\&=&
{\cal T}_i+\sum\limits_{j\ne i}V_i{\cal G}_i{\cal T}_j'
\ee
where we used (\ref{calG}) to write the first line and the definition (\ref{defT}) to derive the last equality.
Using further the identity $V_i {\cal G}_i = {\cal T}_i {\cal G}_0$
the final form reads
\be
{\cal T}'_i(\omega) = {\cal T}_i(\omega) + \sum\limits_{j \neq i} {\cal T}_i(\omega) {\cal G}_0(\omega) {\cal T}'_j(\omega)
\ee
which shows how the single particle channel ${\cal T}$-matrices ${\cal T}_i$ are coupled to yield the auxiliary ${\cal T}$-matrices which add up to the total ${\cal T}$-matrix (\ref{sumT}).

\subsection{Derivation of the Gell-Mann and Goldberger formula}
In the case of only two channels the coupled ${\cal T}$-matrices read
\be
{\cal T}&=&{\cal T}_1'+{\cal T}_2' 
\nonumber\\
{\cal T}_1'&=&{\cal T}_1+{\cal T}_1{\cal G}_0{\cal T}_2'
\nonumber\\
{\cal T}_2'&=&{\cal T}_2+{\cal T}_2{\cal G}_0{\cal T}_1'
\label{Ts}
\ee
and
\be
{\cal T}_1&=&V_1+V_1 {\cal G}_0 {\cal T}_1\nonumber\\
{\cal T}_2&=&V_2+V_2 {\cal G}_0 {\cal T}_2.
\ee 
Introducing ${\cal T}_1'$ into ${\cal T}_2'$  from (\ref{Ts}) leads to
\be
{\cal T}_2'&=&{\cal T}_2(1+{\cal G}_0{\cal T}_1)+{\cal T}_2 {\cal G}_0 {\cal T}_1 {\cal G}_0 {\cal T}_2'
\ee
with the help of which we define
\be
{\cal T}_{ab}\equiv {\cal T}_2' (1+{\cal G}_0 {\cal T}_1)^{-1}&=&{\cal T}_2+{\cal T}_2 {\cal G}_0 {\cal T}_1 {\cal G}_0 {\cal T}_{ab}.
\label{Tab}
\ee
The total ${\cal T}$-matrix from (\ref{Ts}) 
can then be written as
\be
{\cal T}&=&{\cal T}_1+(1+{\cal T}_1{\cal G}_0) {\cal T}_2'
\nonumber\\
&=&{\cal T}_1+(1+{\cal T}_1{\cal G}_0) {\cal T}_{ab} (1+{\cal G}_0 {\cal T}_1).
\label{GMa}
\ee
It remains to show that ${\cal T}_{ab}$ is the ${\cal T}$-matrix  
obeying a Lippman-Schwinger equation with the propagator ${\cal G}_1$. For this purpose we use
(\ref{calG})
\be
-{\cal G}_0&=&-{\cal G}_1+{\cal G}_0V_1{\cal G}_1
\nonumber\\
(1-V_2{\cal G}_1)^{-1}V_2&=&(1-{\cal T}_2{\cal G}_0V_1{\cal G}_1)^{-1} {\cal T}_2
\label{zv}
\ee
where the second line follows from the first by trivial algebra.
Now the right hand side of (\ref{zv}) is just the definition of ${\cal T}_{ab}$ in (\ref{Tab}) using once more ${\cal T}_1{\cal G}_0=V_1 {\cal G}_1$. Therefore we see that from (\ref{zv}) the Lippmann-Schwinger equation (\ref{ab}) for ${\cal T}_{ab}$ follows,
\be
{\cal T}_{ab}=V_2+V_2 {\cal G}_1 {\cal T}_{ab}
\ee
which completes the proof of the Gell-Mann and Goldberger formula (\ref{GG}) or (\ref{GMa}).

\section{Proof of separability}\label{proof}

A separable expansion is always possible for local interactions provided that
the potential range is finite (${\cal V}(r)=0$ for $r>R$). We follow the presentation of Ref. \cite{KPML97} and start from the Lippmann-Schwinger equation with arbitrary energy $\omega$ (called off-shell if $\omega \ne {p^2\over 2m}$),
\be
\ket{\psi}=\ket{p}+{\cal G}_0{\cal V}\ket{\psi}.
\label{ket}
\ee
We
expand the wave function inside the potential range,
\be
\ket{\psi}=\sum\limits_i\ket{B_i}\xi_i,
\label{ket1}
\ee
and multiply (\ref{ket}) by $\bra{B_j}{\cal V}$ from the left to obtain
\be
\sum\limits_i\bra{B_j}{\cal V}\ket{B_i}\xi_i=\bra{B_j}{\cal
V}\ket{p}+\sum\limits_i\bra{B_j}{\cal V}{\cal G}_0{\cal V}\ket{B_i}\xi_i.
\label{matrix}
\ee
Abbreviating
$\mu_{ji}=\bra{B_j}{\cal V}\ket{B_i}$, $\zeta_j=\bra{B_j}{\cal V}\ket{p}$ and
introducing the
form factors
\be
\bra{g_j}=\bra{B_j}{\cal V} \qquad \ket{g_i}={\cal V}\ket{B_i}
\ee
we can write (\ref{matrix}) as a matrix equation 
\be
\sum\limits_i(\mu_{ji}-\bra{g_j}{\cal G}_0\ket{g_i})\xi_i=\zeta_j
\ee
such that the expansion coefficients become
\be
\xi_j=\sum\limits_i\tau_{ij}\zeta_j
\label{excoef}
\ee
with 
$
{\tau}_{ij}^{-1}=\mu_{ij}-\bra{g_i}{\cal G}_0\ket{g_j}.
$
Inserting (\ref{excoef}) into (\ref{ket1}) we have
\be
{\cal V}\ket{\psi}=\sum\limits_{ij}\ket{g_i}\tau_{ij}\bra{g_j}p\rangle
\ee
which shows that the effect of the potential on the scattering wave function 
can be expanded in separable form.

\section{Models for form factors of separable interaction}\label{app}

For different parameterizations of the form factor $g_q$ we can compute the unrestricted two-particle scattering expression of the ${\cal T}$-matrix (\ref{Jinf})
\be
\bra{p_1P}{\cal T}(\omega)\ket{p_2P'}=\lambda \delta_{PP'} {g_{p_1}g_{p_2}\over 1-\lambda J(P,\omega)}
\ee 
which is controlled by
\be
\lambda J(P,\Omega=\omega-{P^2\over 4 m})={\lambda \over 2 \pi^2} \int\limits_0^\infty dq q^2 {g_q^2\over \Omega -{q^2\over m}+i\eta}.
\ee
The phase shift is then given by (\ref{contact}) 
\be
\tan \varphi=\left . {{\rm Im} {\cal T}\over {\rm Re} {\cal T}} \right |_{\Omega={p^2\over m}}=    {\lambda J_{\rm I}\over 1-\lambda J_{\rm R}}
\ee
and the parameter of the potential can be linked to the scattering length $a_0$ and the range of the potential $r_0$ via the small wave-vector expansion
\be
p \cot \varphi={1 \over a_0}+{r_0\over 2 } p^2+....
\label{exp}
\ee
The ${\cal T}$-matrix possesses a pole, the bound state energy $\Omega=-E_b$ is given by
\be
1=\lambda J(-E_b).
\ee
It will be useful to use a dimensionless scaled potential strength $\xi=m \lambda/4 \pi^2 \beta^3$ in the following.

\subsection{Exponential function\label{expo}}

As a first example we choose a form factor 
\be
g_p=\beta^{-2} {\rm e}^{-{p^2/4\beta^2}}
\ee
and obtain for $\Omega>0$ with $y'=\sqrt{{m\Omega/2\beta^2}}$
\be
\lambda J(\Omega)\!=\!-\!{\xi \sqrt{2 \pi}}  \left (\!\!1\!+\!F\left (y'\right )\!+\! i \sqrt{\pi} y'{\rm e}^{-y'^2}\! \right )
\ee
and the Fred-Conte function
\be
F(z)={2 z^2 \over \sqrt{\pi}} \int\limits_0^\infty {dx {{\rm e}^{-x^2}\over x^2-z^2}}.
\ee
The scattering phase shift becomes 
\be
\tan \varphi=-\sqrt{2} \pi \xi {y' {\rm e}^{-y'^2}\over 1+\sqrt{2 \pi} \xi \left (1+F\left (y' \right)\right )}
\ee
from which we obtain the expansion 
\be 
p \cot \varphi=-\beta {\sqrt{2 \pi} \xi+1\over \pi \xi}+\frac{\sqrt{2 \pi} \xi -1}{2\beta \pi \xi} p^2
\ee
and comparing with (\ref{exp}) we can link the potential parameter to the scattering length and the potential range as
\be
a_0&=&-{1 \over \beta}{1\over {1\over \pi\xi}+\sqrt{{2 \over \pi}}}
\nonumber\\
r_0&=&{1 \over \beta} \left (\sqrt{{2\over \pi}}-{1\over \pi \xi}\right).
\label{aoro}
\ee
For negative energies $\Omega=-E_b<0$ we obtain with $y=-i y'=\sqrt{{m E_b / 2 \beta^2}}$
\be
\lambda J(-E_b)=-{\xi \sqrt{2 \pi}}  \left (1-\sqrt{\pi} y {\rm e}^{y^2} {\rm erfc}(y) \right )
\ee
such that the condition for bound states becomes
\be
1+{\xi \sqrt{2 \pi} }  \left (1-\sqrt{\pi} y {\rm e}^{y^2} {\rm erfc}(y) \right )=0
\ee
from which follows that only for $\xi<0$ there can be bound states.
Rewriting it we get
\be
|\xi| \sqrt{2\pi}={1\over 1-\sqrt{\pi} y {\rm e}^{y^2} {\rm erfc}(y)}.
\ee
The right hand side is a monotonically increasing function of $y$ 
starting from $1$ for $y=0$. Therefore, bound states are only possible if
\be
\xi< -{1\over \sqrt{2\pi}}.
\ee

\subsection{Step function}
As a second example, the form factor is assumed to be a step function cut-off in real space $g_r=\Theta(R-r)/4\pi\hbar^2r$ from which we have
\be
g_p={1-\cos{pR/\hbar}\over p^2}.
\ee
For the scattering case, $\Omega=p^2/m>0$, we obtain 
\be
&&\lambda J=\xi \left ({\beta\over p}\right )^3\left [{p\over \beta}+\sin{p\over \beta}(\cos{p\over \beta}-2)-i (1-\cos{p\over \beta})^2\right ]
\nonumber\\&&
\ee
from which the relation to the scattering length and the potential range follows
\be
a_0&=&{1\over 4 \beta} \left (\frac 1 3 +\frac{1}{\xi}\right )^{-1}
\nonumber\\
r_0&=&{2  \over 3 \beta} \left ({11\over 15}-{2\over  \xi}\right ).
\ee

The condition of bound states, for $\Omega=-E_b$, reads
\be
&&\lambda J=-{\xi\over 4 \sqrt{2} y^3} \left (2 \sqrt{2} y +1-({\rm e}^{-\sqrt{2} y}-2)^2\right )
\ee
so that we can obtain bound states if
\be
-{1\over \xi}={2\sqrt{2} y+1-({\rm e}^{-\sqrt{2} y}-2)^2\over 4 \sqrt{2} y^3}
\ee
and since the right hand side is a monotonically decreasing function starting from 1/3 at $y=0$ we obtain the condition for bound states
\be
\xi<-3.
\ee

\subsection{Yamaguchi form factor}

The traditional form factor is 
\be
g_p={1\over \beta^2+p^2}.
\ee
For $\Omega={p^2 \over m}>0$ we obtain 
\be
\lambda J={\xi \beta^2 \pi}{p^2-\beta^2 -2 i p \beta \over 2 (\beta^2+p^2)^2}
\ee
and the scattering phase shift becomes
\be
\cot \varphi={(p^2+\beta^2)^2 \over -\pi \xi p \beta^3}+{p^2-\beta^2 \over 2\beta p}.
\ee
Expanding in $p$ and comparing with (A3) we determine the potential parameters from
\be
a_0=-{2 \over \beta ({2\over \pi \xi}+1) }
\ee
\be
r_0={1 \over \beta}\left (1- {4 \over \xi \pi} \right).
\ee
The case of $\Omega=-E_b$ leads to
\be
\lambda J(-E_b)=-{\pi \xi \over 2 (1+\sqrt{2} y)^2}
\ee
such that we obtain bound states if
\be
\xi=-{2 (1+\sqrt{2} y)^2 \over \pi}.
\ee
Therefore the condition for bound states is
\be
\xi<-{2\over \pi} .
\ee

\section{Integral of (\protect\ref{F})}\label{Fana}

The integral (\ref{F}) can be evaluated analytically. To this end we rewrite (\ref{F}) as
\be
F_\Omega(z)={\Omega \over \pi}\left (\delta_{c,1}+\Omega \int\limits_0^1 d c\right )\int\limits_{-\infty}^{\infty} d y {\sin{cy}\over y-2 z} {1\over n(y)}
\ee
with
\be
n(y)=1+2{\Omega \over y}\sin{y}+2 {\Omega^2\over y^2} (1-\cos{y}).
\label{ny}
\ee
We use the representation of the principal value $1/y=\frac 1 2 (1/(y-i \eta)+1/(y+i \eta))$ and obtain
\be
&&F=\!\left ( \!\Omega\!\delta_{c,1}+\!\Omega^2 \!\int\limits_0^1 d c\!\right ) \!{\rm Re}\!\left ( \!{{\rm e}^{2 i z c}\over n(2 z)}\!+\!2 \sum\limits_\nu{{\rm e}^{i c y_\nu}\over y_\nu\!-\!2 z}{1\over n'(y_\nu)}\!\right )
\nonumber\\
&&=
{\Omega\cos{2 z}\!+\!\Omega^2 {\sin{2 z}\over 2 z}\over n(2z)}
\!+\!2 \sum\limits_\nu
{\Omega {\rm e}^{i y_\nu}\!+\!{\Omega^2 \over i y_\nu} ({\rm e}^{i y_\nu}\!-\!1)\over y_\nu\!-\!2 z}
{1\over n'(y_\nu)}.
\nonumber\\&&
\ee
The values $y_\nu$ represent the complex zeros of $n(y)$. Setting $y_\nu=\mp2 i t_\nu$ these zeros obey
\be
\tanh{t_\nu}=-{t_\nu\over t_\nu+\Omega}.
\label{ti}
\ee
Observing that $1/n(2 z)=A_z^2$ and straightforward algebra using (\ref{ti})
leads to the final representation of the integral as
\be
&&F_\Omega(z)=\Omega  (\cos{2 z}\!+\!\Omega {\sin{2 z}\over 2 z})A_z^2
\!+\!2 \sum\limits_\nu {t_\nu\over t_\nu\!+\!i z} {2 t_\nu\!+\!\Omega\over 1\!+\!\Omega \!+\!2 t_\nu}
.
\nonumber\\&&
\label{Fint}
\ee
The special value required for (\ref{bcond})
reads
\be
&&F_\Omega(0)={\Omega \over \Omega+1}\!+\!2 \sum\limits_\nu  {2 t_\nu+\Omega\over 1+\Omega +2 t_\nu}
\label{Fint0}
\ee
where for $\Omega>0$ the equation (\ref{ti}) has no solution and therefore the last sum in (\ref{Fint}) and (\ref{Fint0}) is absent.

Another straightforward rewriting yields the relation
\be
F_\Omega(ib)-F_\Omega(0)=-b+{2b^2 \over \pi}\int\limits_0^\infty {d x \over x^2+b^2}A_x^2
\ee
which proves $F_\Omega(ib)+b\ge F_\Omega(0)$ used in the discussion following (\ref{bcond1}) where $b=R \sqrt{m E_b}$

\section{Other occurring integrals}\label{integrals}

Here we calculate the integrals appearing in chapter~\protect\ref{finite}.

\subsection{$G_{n_1n_2}(0)$}
The first integral we consider is
\be
&&G_{n_1n_2}(0)\!=\!\int\! d {\bf r} G_{n_1n_2}({\bf r})\!=\!\int\! {d {\bf q}\over (2 \pi)^3} g_q \Psi_{{\bf k}_1}({\bf q})\Psi_{{\bf k}_2}(-{\bf q}).\nonumber\\&&
\label{e1}
\ee
It is convenient here to use the wave function in momentum space. We obtain for the cavity model (\ref{opa})
\be
&&\chi_k(q)\!=\!-4 \pi c R \!\left (\!{ \Omega \sin{(q R)}\sin{(x\!+\!\phi)}\over q^2 R^2\!-\!x^2}\!-\!{\pi \cos{\phi}\over 2} \delta(q R\!-\!x) \! \right )\nonumber\\
&&
\ee
with $x= k R$ or for the three-dimensional wave function
\be
\Psi_k({\bf q})&=&-c R^2\,\bigg[{ \over } { 4 \pi\Omega \sin{(q R)}\sin{(x+\phi)}\over x(q^2 R^2-x^2)}\nonumber\\
&&-x (2 \pi)^3 \delta({\bf q} R-{\bf x}) \cos{\phi}  \bigg] .
\nonumber\\&&
\label{psik}
\ee
Changing variables $y=q R$ in (\ref{e1}) we obtain
\be
&&G_{n_1n_2}(0)={8 R c^2\over \beta^2} \tilde G_0({\bf x}_1,{\bf x}_2)
\ee
and using the exponential form factor of appendix~\ref{expo}, we obtain with $\gamma=2 R \beta$
\be
&&\tilde G_0({\bf x}_1,{\bf x}_2)\!=\!\!\int\limits_0^\infty \!\!d y{\rm e}^{-{y^2/\gamma^2}} {\Omega^2\sin^2{y}\sin{(x_1\!+\!\phi_1)}\sin{(x_2\!+\!\phi_2)}\over (y^2\!-\!x_1^2)(y^2\!-\!x_2^2)}
\nonumber\\&&\qquad-
{\Omega \pi \over 2(x_1^2-x_2^2)}
\left ({\rm e}^{-{x_1^2/\gamma^2}}\sin{x_1} \sin{(x_2+\phi_2)}\cos{\phi_1}
\right .
\nonumber\\&&
\qquad\qquad\qquad\quad\left . \quad +{\rm e}^{-{x_2^2/\gamma^2}}\sin{x_2} \sin{(x_1+\phi_1)}\cos{\phi_2} \right )\nonumber\\
&&\qquad+\pi^3 x_1^2\delta({\bf x_1}+{\bf x_2})\cos{\phi_1}^2{\rm e}^{-{x_1^2/\gamma^2}}.
\label{gt0}
\ee
The remaining integral can be written as
\be
\int\limits_0^\infty  {{\rm e}^{-y^2/\gamma^2}\sin{y}^2\over (y^2-x_1^2)(y^2-x_2^2)}=\left \{\begin{array}{ll}{I_\gamma(x_1^2)-I_\gamma(x_2^2)\over x_1^2-x_2^2} & x_1\ne x_2 \cr &\cr
{1 \over 2 x_1 }{\partial \over \partial x_1} I_\gamma(x_1^2) &x_1=x_2
\end{array}\right . 
\ee
with
\be
I_\gamma(x^2)&=&\int\limits_0^\infty d y {{\rm e}^{-y^2/\gamma^2} \sin{y}^2\over y^2-x^2}={1\over 4 x} \int\limits_{-\infty}^\infty d t {\rm e}^{-t^2} {1-\cos{2 \gamma t}\over t-x/\gamma }
\nonumber\\
&=&{1 \over 4 x} (-\pi {\rm Im} \,W(x/\gamma)-f(2 \gamma,x/\gamma)).
\ee
Here we used the standard integral representation of the complex error function 
for $W(z)={\rm e}^{-z^2} {\rm erfc}{(-i z)}$ and abbreviated
\be
f(a,b)=\int\limits_{-\infty}^\infty d t{{\rm e}^{-t^2} \cos{(a t)}\over t-b}.
\label{fab}
\ee
This integral can be solved analytically. To this end we observe that
\be
f(0,b)&=&-\pi {\rm Im} \, W(b)\\ 
{\partial \over \partial a} f(a,b)|_{a=0}&=&0
\label{bcfab}
\ee
and find a differential equation from (\ref{fab})
\be
\partial_a^2 f(a,b)+b^2 f(a,b)=-b\sqrt{\pi} {\rm e}^{-a^2/4}.
\ee
This linear inhomogeneous differential equation with constant coefficients can be solved straightforwardly and with the boundary conditions (\ref{bcfab}) we obtain finally
\be
f(a,b)= \pi {\rm e}^{-b^2} {\rm Im}\,  {\rm e}^{-i a b } {\rm erf}{({a\over 2}-i b)}.
\ee

\subsection{$\int dp G_{n_1n_2}(p)$}

The second integral used in (\ref{inte}) is conveniently expressed by the wave functions in spatial representation
\be
\int {d {\bf p}\over (2 \pi)^3} G_{n_1n_2}(p)=\int d {\bf r} g_r \Psi_{{\bf k}_1} ({\bf r}/2) \Psi_{{\bf k}_2}(-{\bf r}/2)
\label{intgp}
\ee
where we use the Fourier transform of (\ref{psik})
\be
&&\Psi_{\bf k}({\bf r})=c k {\rm e}^{i{\bf k r}}\cos{\phi} +{c\sin{\phi} \over r}\!\left \{\!
\begin{array}{ll}
 \cot x \sin{(k r)}& r<R\cr
\cos{(k r)} & r>R
\end{array}\right ..
\nonumber\\&&
\label{wavep}
\ee
For the exponential form factor of appendix~\ref{expo} the integrals in (\ref{intgp}) becomes
\be
\int {d {\bf p}\over (2 \pi)^3} G_{n_1n_2}(p)={32 R \beta c^2\over \sqrt{\pi}} {\bar G}({x}_1,{x}_2).
\ee

The integral (\ref{gt0}) contains a term $\sim \delta({\bf x_1}+{\bf x_2})$ and will be multiplied with (\ref{intgp}) as part of the integrand in (\ref{js}). Therefore special care is needed to evaluate (\ref{intgp}) for the diagonal case ${\bf x_1}=-{\bf x_2}$. 
We split therefore ${\bar G}({x}_1,{x}_2)={\bar G}'({x}_1,{x}_2)+{\bar G}''({x}_1)$ where we use for the first term the radial component of (\ref{wavep}) and summarize in the second part the difference of the angular integral over the wave functions and the radial one. This difference vanishes except for the case ${\bf x_1}=-{\bf x_2}$.

Changing variables $y=r/R$ in (\ref{intgp}) we obtain for the regular part ${\bar G}'$
\be
&&{\bar G}'(x_1,x_2)=A_{x_1}^2A_{x_2}^2\int\limits_0^1 dy {\rm e}^{-y^2 \gamma^2} \sin{x_1 y}\sin{x_2 y}\nonumber\\
&&\quad+\int\limits_1^\infty dy {\rm e}^{-y^2 \gamma^2} \sin{(x_1 y+\phi_1)}\sin{(x_2 y+\phi_2)}
\ee
which are trivial integrals. 
The irregular part ${\bar G}''$ is now the difference between the angular averaging of the vectorial part of the wave functions $\sim  k_1 r \sin{k_1 r}$ and the expression of the radial part of the wave function $\sim \sin^2{k_1 r}$ since the latter is already absorbed in the regular part. We obtain
\be
&&{\bar G}''(x_1)=\cos^2{\phi_1} \int\limits_0^\infty dy {\rm e}^{-y^2 \gamma^2} (x_1 y \sin{(2 x_1 y)}-(\sin{x_1 y})^2)
\nonumber\\&&
\ee
which again is a trivial integral.

\bibliography{kmsr,kmsr1,kmsr2,kmsr3,kmsr4,kmsr5,kmsr6,kmsr7,delay2,spin,refer,delay3,gdr,chaos,sem3,sem1,sem2}

\begin{thebibliography}{28}
\bibitem[{aoc()}]{aocomm}
\bibinfo{note}{We follow here the convention of Ref. \protect\cite{GW64} which
  has an opposite sign compared to Ref. \protect\cite{F94}}.
\bibitem[{sig()}]{signcomm}
\bibinfo{note}{Please note that for positive scattering lengths there is an
  even number of bound states and for negative scattering lengths there is an
  odd number \protect \cite{GW64}. Only for contact potentials the sign of the
  coupling strength follows the sign of the scattering length}.
\end{thebibliography}

\end{document}